\documentclass{Interspeech2024}

\PassOptionsToPackage{hyphens}{url}
\usepackage{hyperref}
\usepackage[table]{xcolor}
\definecolor{color0}{rgb}{0.4, 0.5, 0.4}    
\definecolor{color3}{rgb}{0.8, 0.7, 1}  
\definecolor{color2}{rgb}{1, 0.7, 0.7}     
\definecolor{color5}{rgb}{0.7, 1, 0.7}   
\definecolor{color1}{rgb}{1, 1, 0.7}    
\usepackage{todonotes}
\usepackage{graphicx,tabularx}
\usepackage{booktabs} 

\usepackage{amssymb,amsthm,amsmath}
\usepackage{pgfplots}
\usepackage{booktabs}

\usepackage{cleveref}
\pgfplotsset{compat=1.16}

\newcommand{\circled}[1]{%
  \tikz[baseline=(char.base)]{
    \node[shape=circle,draw,inner sep=0.8pt,minimum size=1.0em,text centered] (char) {#1};}
}

\newcounter{idcounter}
\newcommand{\id}[1]{%
  \ifcsdef{idname#1}{%
    \circled{\csuse{idname#1}}
  }{%
    \stepcounter{idcounter}
    \csxdef{idname#1}{\arabic{idcounter}}
    \circled{\arabic{idcounter}}
  }%
}

\interspeechcameraready

\title{A New Approach to Voice Authenticity}

\name[affiliation={1}]{Nicolas M.}{Müller}
\name[affiliation={2}]{Piotr}{Kawa}
\name[affiliation={3}]{Shen}{Hu}
\name[affiliation={4}]{Matthias}{Neu}
\name[affiliation={5}]{Jennifer}{Williams}
\name[affiliation={1}]{Philip}{Sperl}
\name[affiliation={1}]{Konstantin}{Böttinger}

\address{
  $^1$Fraunhofer AISEC 
  $^2$Wrocław University of Science and Technology 
  $^3$Technical University of Munich 
  $^4$Bundesamt für Sicherheit in der Informationstechnik 
  $^5$University of Southampton 
  }
\email{nicolas.mueller@aisec.fraunhofer.de}


\begin{document}

\maketitle

\begin{abstract}

Voice faking, driven primarily by recent advances in text-to-speech (TTS) synthesis technology, poses significant societal challenges.
Currently, the prevailing assumption is that unaltered human speech can be considered genuine, while fake speech comes from TTS synthesis. 
We argue that this binary distinction is oversimplified. 
For instance, altered playback speeds can be used for malicious purposes, like in the `Drunken Nancy Pelosi' incident.
Similarly, editing of audio clips can be done ethically, e.g. for brevity or summarization in news reporting or podcasts, but editing can also create misleading narratives. 
In this paper, we propose a conceptual shift away from the binary paradigm of audio being either `fake' or `real'. Instead, our focus is on pinpointing `voice edits', which encompass traditional modifications like filters and cuts, as well as TTS synthesis and Voice Conversion (VC) systems.
We delineate 6 categories and curate a new challenge dataset rooted in the M-AILABS corpus, for which we present baseline detection systems.
And most importantly, we argue that merely categorizing audio as fake or real is a dangerous over-simplification that will fail to move the field of speech technology forward. 

\end{abstract}



\section{Introduction}
The rapidly advancing field of machine learning has significantly enhanced the quality and computability of text-to-speech (TTS) synthesis technology, opening the door to a myriad of beneficial applications~\cite{parrotron,Respeech99,tts-audiobooks,siri}.
However, that same forward progress also brings forth serious threats to our understanding and perception of authentic speech, including the creation of deepfakes aimed at deceiving the human ear. The result has become more clear in recent years with the spread of misinformation, fake news, slander, fraud, AI-mediated pornography, and deceptive calls~\cite{audio-scam, audio-scam-2, audio-fake-news}.
Speech synthesis technology becomes particularly perilous when used to trick automatic speaker verification and biometric identification systems - a practice known as `spoofing'~\cite{vice-breaking-bank}.

The anti-spoofing community has responded by establishing challenge datasets and spoofing detection algorithm benchmarks such as ASVspoof 2015, 2017, 2019~\cite{todisco2019asvspoof} and 2021~\cite{asvspoof2021}, pouring extensive efforts into differentiating between genuine (`bona-fide') and counterfeit (`spoof') speech samples.
While these datasets, algorithms, and evaluation metrics are adequate within the scope of anti-spoofing protection for speaker verification, they fall short in addressing broader societal challenges posed by technologies used for making deepfakes.
Particularly, the binary idea that all TTS/VC-generated synthetic speech is inherently deceptive (and fake), while all other content is benign (and genuine), is an oversimplification.
Despite this false equivalence, it is a viewpoint widely held both inside and outside of the speech technology community and among voice authenticity researchers~\cite{todisco2019asvspoof,asvspoof2021,in-the-wild,add2022}.

The reality of assessing fake and genuine audio is complex and context-dependent. For instance, altering playback speeds - a technique maliciously used in scenarios like the `Drunken Nancy Pelosi'~\cite{drunk_pelosi} incident - can also serve benign purposes, such as in language learning tools or assistance for the hearing-impaired. 
Similarly, even the removal or reordering of words can fabricate misleading narratives~\cite{FoxNewse71} that use real human speech, yet these same editing techniques are valid in concise, legitimate formats such as news broadcasts or podcasts. 
Finally, audio can be edited through equalization to subtly alter the quality of a politician's voice by adjusting bass and treble levels, as well as overall pitch. Equalization adjustments can affect how sincere or confident a politician sounds. 
Even unintentional manipulation can degrade speech sound quality, thereby negatively influencing public perception~\cite{speech_edits_pol}, which has already been observed in political campaigns~\cite{TVPcelow97}.


Furthermore, not all synthetic speech warrants skepticism. Consider the speech synthesis used by the late physicist Stephen Hawking for accessibility purposes.
Likewise, the `Google Parrotron'~\cite{parrotron} project aims to help people with partial speech impairments, using the same TTS synthesis techniques that enable spoofs and deepfakes.
Lastly, in the field of politics, TTS synthesis has been used both ethically and unethically in automated calls.
One the one hand, it has been used to impersonate US President Biden and encourage people to skip the 2024 primary election in the state of New Hampshire~\cite{robocall_biden}. On the other hand, the US Democratic Party has used it to connect with new potential voters~\cite{robocall_ashley}.
This underscores the necessity for a nuanced understanding of `synthetic' and `authentic' in audio content—a delineation that is not merely black and white but incorporates the myriad shades of intention and context.

\textbf{Contribution.} Our paper introduces a structured approach and new way of thinking about challenges for handling voice-edited audio. 
We introduce a new dataset that reflects this paradigm shift. We also propose several baseline machine learning models capable of identifying and classifying audio modifications in the dataset, including the nature of the edit and its location in the time domain. 
Our evaluation confirms that models are effective at accurately detecting a variety of audio modifications and edits. 
Using this dataset as the basis, we call for a fundamental shift in voice authenticity research efforts, away from the simplistic real/fake classification and toward a more nuanced approach.


\section{Related Work}

\subsection{Detection of Audio Modifications}
The domain of audio modification detection encompasses various challenges, including the identification of double compression, codec recognition, and the detection of copy-move or splicing operations, which involve the deletion or insertion of audio segments~\cite{double-mp3,audio-tampering-survey,audio-tampering-survey-2}. 
Efforts in this area include the analysis of electric network frequency (ENF) signals~\cite{esquef2014edit}, as well as investigations into microphone and acoustic characteristics~\cite{audio-tampering-survey-2}. 
Despite these endeavors, the literature remains sparse, with no existing work providing as detailed a classification of vocal alterations or as current an overview of neural network-based detection methodologies as presented in this study.




\subsection{Voice Authenticity Attacks}
Voice authenticity attacks are commonly categorized into three primary scenarios: physical access spoofing, logical access spoofing, and deepfakes.
This taxonomy is reflected in the datasets developed by the research community.
Physical access spoofing targets automatic speaker verification (ASV) systems with replay attacks, where utterances are recorded and replayed under varied acoustic conditions~\cite{todisco2019asvspoof,asvspoof2021,asvspoof-2017,idiap-voicepa}. 
Logical access spoofing entails remote attacks on ASV systems, such as biometric identification in telephone call. 
Corresponding techniques include the injection of spoofed audio, created using text-to-speech and voice conversion techniques, into the communication channel~\cite{asvspoof2021,idiap-avspoof,asvspoof-2015}. 
Deepfake technology aims to generate artificial speech that can deceive human listeners, focusing on disinformation spread via social media, using similar techniques as logical access spoofing but targeting humans instead of voice biometric systems~\cite{asvspoof2021,add2022,add2023,cfad}.

\subsection{Deeepfake and Spoofing Detection}
The development of anti-spoofing and deepfake detection algorithms primarily relies on deep neural networks, diverging into two main branches based on the nature of the input data. 
The first branch utilizes raw audio waveforms, requiring no prior transformations~\cite{tak2021EndtoEnd,ge2021raw,aasist}, while the second focuses on transformed audio signals to highlight features indicative of spoofed material~\cite{wang2021Comparative, muller2023complex}. 
Self-supervised learning models and embeddings from other audio processing architectures have also been proposed~\cite{zhang2021fake,ssl_antispoof,whisper-df}.
Challenges in the detection of voice spoofs and deepfakes include ensuring the generalization of trained methods. 
This is currently addressed by incorporating a wide array of generation methods, codecs, and audio quality within the training datasets. 
Other focal points include partial spoofs, multi-modal deepfakes, synthesized singing voices detection, real-world deepfake utterances, and language diversity~\cite{in-the-wild,cfad,had,partial-spoof,fakeavceleb,sing-fake,mlaad}. 
Despite the extensive taxonomy, existing classifications do not consider traditional attack vectors such as slicing and pitch alteration. 
There is no comprehensive database or benchmark addressing both neural and `traditional' audio manipulations.

\section{Voice Edit Categories}

In table~\ref{tab:veds}, we identify 6 overarching categories of voice edits that contribute to voice authenticity. 
Each category is based on a different type of audio modification which can affect the perception of voice and voice quality. 
From these high-level categories, we present~21 unique voice edits (encircled in numerals).

\begin{table}[!htb]
\centering
\caption{Voice Edit Categories}
\label{tab:veds}
\begin{tabularx}{\columnwidth}{X}
\toprule
\textbf{Source Origin} \\[-0.8em] 
\begin{itemize}
    \itemsep0em
    \item \textit{Original}: Unaltered human speech \id{original_voice}.
    \item \textit{Synthesized}: Speech generated using text-to-speech synthesis algorithms \id{text_to_speech}.
    \item \textit{Converted}: Speech generated by voice conversion \id{voice_conversion}.
\end{itemize} \\
\midrule
\textbf{Temporal Edits} \\[-0.8em]
\begin{itemize}
    \itemsep0em
    \item \textit{Concatenation/Trimming}: Inserting or removing portions of the audio \id{insert_or_cut_audio}.
    \item \textit{Mixing}: Merging two separate audio files together \id{mixing}.
\end{itemize} \\
\midrule
\textbf{Modulation and Effects} \\[-0.8em]
\begin{itemize}
    \itemsep0em
    \item \textit{Pitch Shifting}: Increase or decrease the pitch without altering the speed \id{pitch_up}, \id{pitch_down}.
    \item \textit{Time Stretching}: Slow down or speed up the pace without altering the pitch \id{speed_slower}, \id{speed_faster}.
\end{itemize} \\
\midrule
\textbf{Encoding and Compression} \\[-0.8em]
\begin{itemize}
    \itemsep0em
    \item \textit{Lossless Encodings}: Formats like WAV, FLAC are considered the `default' encoding.
    \item \textit{Lossy Encodings}: Formats like MP3 or AAC \id{mp3_compression}, \id{aac_compression}.
    \item \textit{Telephony Encodings}: Specific to voice communications, e.g., alaw \id{alaw_encoding}, ulaw \id{ulaw_encoding}.
\end{itemize} \\
\midrule
\textbf{Frequency and Spectral Edits} \\[-0.8em]
\begin{itemize}
    \itemsep0em
    \item \textit{Low/High Pass Filters}: Filters designed to allow only low or high frequency ranges to pass through \id{low_pass_filter}, \id{high_pass_filter}.
    \item \textit{Equalization}: Adjusting the balance between frequency components \id{equalization}.
    \item \textit{Autotune}: Correcting pitch in vocal performances \id{auto_tune}.
\end{itemize} \\
\midrule
\textbf{Spatial and Environmental Edits} \\[-0.8em]
\begin{itemize}
    \itemsep0em
    \item \textit{Room Impulse Response}: The modification of audio to simulate different room acoustics \id{room_impulse}.
    \item \textit{Reverb}: Adding reflections \id{reverb}.
    \item \textit{Overlay}: Mix 
    with background noise \id{overlay_background}.
    \item \textit{Noise Cancellation}: Removing background noises \id{noise_reduce}.
\end{itemize} \\
\end{tabularx}
\end{table}

\section{Data}
\subsection{Human speech baseline}
We introduce a framework designed for the efficient, real-time generation of the those 21voice edits. 
To this end, we use the M-AILABS Speech Dataset~\cite{mailabs}. 
This dataset comprises eight languages: English, French, German, Italian, Polish, Russian, Spanish, and Ukrainian. 
It is based on LibriVox and Project Gutenberg, encompassing almost 1000 hours of audio across 493,900 baseline human utterances \id{original_voice}, and sampled at 16kHz.


\subsection{TTS and VC data creation}
In order to generate corresponding text-to-speech data \id{text_to_speech}, we use the MLAAD dataset~\cite{mlaad}, which also uses the M-AILABS dataset as a reference point, and re-synthesizes the audio files using 52 different text-to-speech algorithms.
Since MLAAD does not feature any voice-conversion \id{voice_conversion}, we create the corresponding data ourselves.
To this end, we use FreeVC~\cite{freevc} and Phoneme Hallucinator~\cite{shan2023phoneme} to create 1000 converted audio samples for each language of the M-AILABS dataset and each Voice Conversion method. The target speaker for each source speaker is a randomly chosen speaker of the same language.


\subsection{On-the-Fly Modifications}

The remaining modifications \id{insert_or_cut_audio} to \id{noise_reduce} offer a computational efficiency significantly surpassing that of neural Text-to-Speech systems.
This enables their real-time execution during both training and testing phases; thus serving as on-the-fly data augmentations. 
These alterations are generally straightforward; for instance, the `Concatenation/Trimming' operation \id{insert_or_cut_audio} entails choosing two arbitrary points within audio files to concatenate together. 
Similarly, other modifications such as altering speech or pitch are readily achievable through efficient utilization of the \emph{librosa}, \emph{ffmpeg} 
 or \emph{sox} libraries.


All modifications necessitate the specification of hyperparameters. 
For instance, adjustments like increasing pitch \id{pitch_up} or decreasing pitch \id{pitch_down} demand a specification of the semi-tone interval for the shift. 
We randomly determine this value within a pre-established spectrum, in this case, between 1 and 12 semi-tones.
The modification is then applied and the file labelled accordingly.
A similar approach applies to other voice modifications, where suitable hyperparameters are not fixed, but chosen from a range of suitable values.
Thus, this process results in a dynamically created set of audio files.

\begin{table*}[t]
    \centering
    \caption{Model detection performance relative to varying input durations. Color coding denotes optimal performance within the same input lengths: yellow for the best results at 0.35 seconds, red for 1.2 seconds, and purple for 4.05 seconds. Results are presented as mean ± standard deviation.}
    \label{tab:res}
    \resizebox{.999999\textwidth}{!}{%
        \begin{tabular}{llll|lll|lll|l}
\toprule
Model Name &                   ComplexNet &                    &                    &                          LCNN &                          &                          &                     RawNet2 &     &     &                     SSL W2V2 \\
Input Length                              &                         0.35s &                           1.2s &                          4.05s &                          0.35s &                           1.2s &                          4.05s &                        0.35s &         1.2s &        4.05s &                         4.05s \\
\midrule
Epoch Time (min)                             &                      4.0±0.8 &                       5.6±0.1 &                      15.9±0.1 &                       4.3±0.1 &                       5.0±0.3 &                       7.6±2.0 &                     2.5±0.1 &     2.9±0.5 &     5.5±0.5 &                     34.0±0.3 \\
Test Accuracy                                &                     51.3±1.1 &   \cellcolor{color2} 72.1±1.8 &                      75.4±0.0 &   \cellcolor{color1} 55.0±2.2 &                      70.2±1.3 &                      69.0±6.5 &                    10.9±1.5 &    51.8±0.9 &    52.7±0.4 &  \cellcolor{color3} 80.8±5.4 \\
\midrule
\textbf{F1 Scores} &&&&&& \\
Original Voice \id{original_voice}           &                      2.2±1.7 &  \cellcolor{color2} 16.6±13.1 &                      36.8±3.2 &   \cellcolor{color1} 7.4±10.4 &                       4.4±3.1 &                     22.2±22.2 &                     0.0±0.0 &     3.9±5.6 &    16.5±3.1 &  \cellcolor{color3} 48.7±9.9 \\
Text To Speech \id{text_to_speech}           &                    35.5±25.2 &   \cellcolor{color2} 77.6±3.0 &                      90.9±1.3 &   \cellcolor{color1} 67.8±3.5 &   \cellcolor{color2} 78.0±7.4 &                     72.8±22.6 &                     0.0±0.0 &    54.4±4.8 &    58.0±8.8 &  \cellcolor{color3} 97.8±1.0 \\
Voice Cloning \id{voice_conversion}             &                    75.3±11.8 &   \cellcolor{color2} 95.9±0.8 &                      98.4±0.5 &  \cellcolor{color1} 81.3±12.4 &                      94.6±4.0 &                      89.9±9.8 &                    46.9±3.2 &   81.6±14.1 &   74.9±27.9 &  \cellcolor{color3} 99.8±0.2 \\
Concatenation/Trimming \id{insert_or_cut_audio} &   \cellcolor{color1} 0.0±0.0 &   \cellcolor{color2} 16.1±8.8 &                      12.2±8.1 &    \cellcolor{color1} 0.0±0.0 &                      15.9±3.2 &                       5.7±8.1 &  \cellcolor{color1} 0.0±0.0 &     8.0±4.4 &     1.1±0.1 &  \cellcolor{color3} 92.8±2.4 \\
Mixing \id{mixing}                           &                     22.8±3.6 &   \cellcolor{color2} 78.6±1.8 &                      92.5±2.2 &   \cellcolor{color1} 25.9±2.3 &                      62.0±6.0 &                      83.4±1.7 &                     0.0±0.0 &   40.1±12.4 &   56.6±25.1 &  \cellcolor{color3} 97.8±1.1 \\
Pitch Up \id{pitch_up}                       &  \cellcolor{color1} 81.1±3.0 &   \cellcolor{color2} 94.3±1.2 &                      96.7±0.7 &                      78.0±1.7 &                      89.7±1.5 &                     86.6±13.4 &                     0.0±0.0 &    74.0±3.5 &    81.9±1.7 &  \cellcolor{color3} 98.0±1.0 \\
Pitch Down \id{pitch_down}                   &                     70.6±7.0 &                      94.8±3.8 &                      97.9±1.9 &   \cellcolor{color1} 90.7±1.2 &   \cellcolor{color2} 97.8±1.1 &   \cellcolor{color3} 99.3±0.6 &                     0.0±0.0 &    77.3±6.3 &   75.7±16.2 &                     98.3±1.4 \\
Speed Slower \id{speed_slower}               &  \cellcolor{color1} 70.9±3.0 &   \cellcolor{color2} 75.8±4.1 &                     46.1±32.5 &                     34.5±18.0 &                     29.4±26.5 &                      53.9±6.5 &                     0.0±0.0 &    40.1±1.8 &   48.7±16.6 &  \cellcolor{color3} 83.3±3.3 \\
Speed Faster \id{speed_faster}               &  \cellcolor{color1} 73.5±5.7 &   \cellcolor{color2} 79.1±2.8 &                     50.1±21.9 &                      62.1±6.8 &                      64.8±2.7 &                     43.1±26.8 &                     0.0±0.0 &    56.4±4.0 &   32.5±13.1 &  \cellcolor{color3} 83.5±0.9 \\
Mp3 Compression \id{mp3_compression}         &                     54.9±1.9 &                      84.4±5.4 &                      90.3±4.5 &   \cellcolor{color1} 86.4±1.1 &   \cellcolor{color2} 96.8±1.4 &   \cellcolor{color3} 98.1±2.4 &                     1.3±1.8 &    15.2±4.8 &    25.3±6.9 &                    79.7±19.5 \\
Aac Compression \id{aac_compression}         &                     91.2±2.9 &   \cellcolor{color2} 91.6±1.3 &   \cellcolor{color3} 85.4±2.1 &   \cellcolor{color1} 92.8±3.9 &                      84.8±1.7 &                     54.2±16.2 &                     2.7±3.8 &    83.8±3.6 &     6.9±0.9 &                     81.9±1.5 \\
Alaw Encoding \id{alaw_encoding}             &                    15.9±22.5 &                     12.1±17.1 &  \cellcolor{color3} 45.0±20.4 &   \cellcolor{color1} 41.5±1.1 &  \cellcolor{color2} 40.2±17.0 &                     35.4±19.5 &                     0.0±0.0 &     5.7±2.1 &     2.8±4.0 &                    32.6±41.8 \\
Ulaw Encoding \id{ulaw_encoding}             &                    14.4±13.2 &  \cellcolor{color2} 47.0±12.0 &                     24.6±33.2 &   \cellcolor{color1} 24.7±2.2 &                     44.4±18.0 &  \cellcolor{color3} 53.8±12.3 &                     0.0±0.0 &     0.4±0.6 &     0.0±0.0 &                    12.4±17.5 \\
High Pass Filter \id{high_pass_filter}       &  \cellcolor{color1} 88.6±0.6 &   \cellcolor{color2} 93.6±0.5 &   \cellcolor{color3} 96.1±0.6 &                      48.3±3.7 &                      91.6±1.9 &                      91.6±2.1 &                    14.0±3.6 &    92.5±3.4 &    92.6±1.9 &  \cellcolor{color3} 96.4±0.6 \\
Low Pass Filter \id{low_pass_filter}         &                     22.4±5.5 &                     58.6±12.7 &                      69.3±4.4 &   \cellcolor{color1} 59.9±0.6 &   \cellcolor{color2} 77.6±1.0 &  \cellcolor{color3} 70.0±21.5 &                     0.0±0.0 &    16.7±5.7 &   13.2±14.6 &                    50.8±44.8 \\
Equalization \id{equalization}               &  \cellcolor{color1} 18.2±6.5 &                     17.9±25.3 &  \cellcolor{color3} 19.4±27.5 &                      17.0±9.8 &   \cellcolor{color2} 43.9±3.3 &                      7.4±10.5 &                     5.2±7.4 &     7.1±4.2 &     0.0±0.0 &                      0.0±0.0 \\
Auto Tune \id{auto_tune}                     &                    42.4±19.5 &   \cellcolor{color2} 78.9±3.6 &                      95.1±1.5 &   \cellcolor{color1} 45.8±1.5 &                      77.6±2.1 &                      80.1±8.1 &                     0.0±0.0 &    47.2±1.9 &    73.8±8.4 &  \cellcolor{color3} 98.3±0.5 \\
Room Impulse \id{room_impulse}               &  \cellcolor{color1} 91.7±0.5 &   \cellcolor{color2} 97.9±0.9 &   \cellcolor{color3} 98.7±0.4 &                      84.9±2.4 &                      90.4±3.0 &                      84.1±8.4 &                    42.6±1.1 &    92.6±1.5 &    88.0±2.5 &  \cellcolor{color3} 99.4±0.1 \\
Reverb \id{reverb}                           &  \cellcolor{color1} 72.2±4.3 &   \cellcolor{color2} 96.8±1.4 &   \cellcolor{color3} 99.3±0.3 &                     45.7±14.8 &                      86.3±1.5 &                      87.0±4.7 &                     0.0±0.0 &    87.3±4.6 &    84.1±8.2 &  \cellcolor{color3} 99.9±0.2 \\
Overlay Background \id{overlay_background}   &  \cellcolor{color1} 80.5±0.9 &   \cellcolor{color2} 96.4±0.6 &                      98.3±0.0 &                      75.6±6.8 &                      91.8±0.1 &                      95.5±0.2 &                     0.0±0.0 &    86.3±9.3 &    85.6±8.0 &  \cellcolor{color3} 99.5±0.4 \\
Noise Reduce \id{noise_reduce}               &                    39.8±17.8 &   \cellcolor{color2} 82.4±4.4 &   \cellcolor{color3} 99.7±0.0 &   \cellcolor{color1} 54.0±8.2 &                      62.5±1.5 &   \cellcolor{color3} 99.4±0.9 &                     0.0±0.0 &    70.5±1.2 &    92.1±0.3 &  \cellcolor{color3} 99.3±0.3 \\
\bottomrule
\end{tabular}
    }
\end{table*}

\section{Experiments}
Given such a data framework comprising 21voice edits, we evaluate the detection performance of several neural architectures from related work.
We follow standard procedure:
The M-AILABS is split into train and test utterances (90\% and 10\%, respectively), where each audio-file is transformed on-the-fly and used to train or evaluate the models. 
We use batch sizes of 16, the Adam optimizer with a learning rate of $1e^{-3}$, and train for $500$ epochs.
We employ early stopping when there is no increase in training accuracy of at least $0.5\%$ within $5$ epochs.

\subsection{Model Description}
We use four machine learning models from the field of audio deepfake detection and anti-spoofing.
These models are emblematic of the most widely utilized architectural types in this domain, namely: 
Model with spectral frontends, both real- and complex-valued, as well as end-to-end models, either with or without pre-training. 

\begin{itemize}
    \item \textbf{Light Convolutional Neural Network (LCNN)}~\cite{wang2021Comparative}: 
    LCNNs are distinguished for their robust performance in anti-spoofing tasks, employing a mel-spectrogram or CQT frontend, where only the magnitude of the time-frequency spectrogram is analyzed. This approach can be fast and reliable, but incurs some information loss because the phase information is discarded.
    \item \textbf{ComplexNet}~\cite{muller2023complex}: 
    Unlike traditional methods that often overlook phase information, ComplexNet efficiently processes CQT-features \cite{brown1991calculation} extracted directly from raw audio waveforms without discarding the phase data. This method ensures no loss of information during feature extraction, requiring however the use of complex-valued neural layers to handle the resulting complex-valued time-frequency spectrogram adequately.
    \item \textbf{RawNet2}~\cite{tak2021EndtoEnd}: 
    As a one of the most popular models in voice anti-spoofing, RawNet2 directly processes the raw audio waveform using sinc-layers, bypassing the need for a dedicated spectral frontend. Despite its notable success on ASVspoof, RawNet2 faces challenges in generalizing to out-of-domain data~\cite{in-the-wild}.
    \item \textbf{SSL W2V2}~\cite{ssl_antispoof}: 
    This model capitalizes on unsupervised pre-training via wav-to-vec-2 representations \cite{baevski2020wav2vec}, demonstrating significant potential in cross-domain anti-spoofing scenarios~\cite{mlaad}. However, its computational demands are high, and it requires at least 4.05 seconds of input audio. This requirement limits its utility for fine-grained temporal analysis, albeit potentially enhancing detection capabilities.
\end{itemize}

\subsection{Time Resolution vs. Model Accuracy}
It is crucial to not only determine the type of manipulation an audio file has undergone, but also pinpointing the moment of alteration. 
For instance, in a public debate recording, an adversary might alter the voice of specific individuals to discredit them selectively. 
Thus, in voice edit detection, two primary objectives emerge: 
achieving high detection accuracy and maintaining fine time resolution. 
These objectives often conflict because analyzing a larger audio sample enhances the model's ability to identify voice edits but simultaneously reduces the temporal resolution.

Moreover, some detection models necessitate a minimum audio input length, inherently limiting their capacity for fine time resolution. 
Attempting to artificially extend audio to meet minimum length requirements would itself be considered a form of voice manipulation by concatenation \id{insert_or_cut_audio}, thereby precluding high temporal resolution with these models.

To assess this trade-off, we examine all models - excluding SSL W2V2 - at three levels of time resolution: 
fine ($0.35$s), medium ($1.2$s), and coarse ($4.05$s). This approach involves segmenting the original or modified audio into subsections of these specified lengths, and then supplying the subsections to the neural architecture. 
We choose $0.35$s, because it is the lowest limit allowed by the LCNN model, $4.05$s as the minimum required by SSL W2V2, and $1.2$s as the intermediate value, calculated based on a geometric progression.

\section{Results}
We assess the performance of the trained models with respect to two metrics: overall test accuracy and individual F1 scores for each voice edit. 
The experiments are conducted twice, and the aggregated results, including mean and standard deviation, are presented in Table \ref{tab:res}.
Our evaluation extends beyond mere detection performance; we also examine performance relative to input length, highlighting the best performance for each category and input length. 
We use color coding for clarity: yellow for fine, red for medium, and purple for coarse time resolution.

Our observations are as follows. 
First, as the resolution becomes finer, the detection performance decreases: an overall test accuracy of $80.8\%$ is observed at coarse time resolution, $73.1\%$ at medium resolution, and $55.0\%$ at fine resolution, as shown in Table \ref{tab:res}.
This is expected, since finer time resolution means less data for the model to base its decision on.
Second, we identify a strong trend in detection performance across resolutions:
the SSL W2V2 model excels at coarse resolution, while the LCNN and ComplexNet models perform best at medium and fine resolutions, respectively. 
RawNet2 underperforms across the board. 
Third, not all voice edits pose the same level of detection difficulty. The `Concatenation/Trimming' edit \id{insert_or_cut_audio} is detectable only by the SSL W2V2 model, suggesting that pre-training experience is crucial in this case. 
Many edits are detectable even at fine resolution, whereas some, like `Alaw Encoding' \id{alaw_encoding} and `Equalization' \id{equalization}, are consistently challenging due to minimal audio changes. 
Nevertheless, models like SSL W2V2 demonstrate satisfactory performance when coarse time resolution is adequate.



\section{Discussion and Use Cases}
This section explores several real-world use cases for voice edit detection.
\begin{itemize}
\item \textbf{Fighting Misinformation and Slander:} As observed in the real world, changes in playback speed, equalization, insertion or cutting of audio, as well as TTS and VC recordings can be used to spread misinformation and facilitate fraud~\cite{speech_edits_pol, TVPcelow97, ceo_voice}, making the detection thereof a paramount challenge. 
Relevant edits include `Text To Speech' \id{text_to_speech}, `Voice Conversion' \id{voice_conversion},
`Concatenation/Trimming' \id{insert_or_cut_audio},  `Pitch Up' \id{pitch_up}, `Pitch Down' \id{pitch_down}, `Speed Slower' \id{speed_slower}, `Speed Faster' \id{speed_faster}, `Equalization' \id{equalization}.

    \item \textbf{Legal Authenticity Verification:} Determining the authenticity of audio evidence in court cases to discern between genuine and synthetic or altered voices. 
    Relevant edits include `Text To Speech' \id{text_to_speech}, `Voice Conversion' \id{voice_conversion},  `Concatenation/Trimming' \id{insert_or_cut_audio}, and `Overlay Background Noise' \id{overlay_background_noise}.
    
    \item \textbf{Security and Authentication:} Ensuring voice biometric systems can differentiate between real and synthetic voices for secure access. Critical edits to detect include `Original Voice' \id{original_voice} and `Text To Speech' \id{text_to_speech} and `Voice Conversion' \id{voice_conversion}.

    \item \textbf{Copyright and Piracy Detection:} Identifying unauthorized use or alterations of copyrighted audio content, including compression to evade detection. Relevant edits include encodings to MP3, AAC, Alaw and Ulaw \id{mp3_compression}, \id{aac_compression},
    \id{alaw_encoding}, \id{ulaw_encoding}, as well as `Mixing' \id{mixing} and `Equalization' \id{equalization}.
    
    \item \textbf{Insurance Claim Verification:} Assessing the authenticity of audio evidence in claims, such as verifying the sound environment of an accident scene. Edits of interest: `Room Impulse' \id{room_impulse}, `Reverb' \id{reverb}, `Overlay Background Noise' \id{overlay_background} and `Noise Reduce' \id{noise_reduce}.
    
    \item \textbf{Performance Authenticity:} Ensuring that live or recorded performances in the entertainment industry are free from unauthorized enhancements or effects. Relevant edits: `Auto Tune' \id{auto_tune} and `Equalization' \id{equalization}.
    
    \item \textbf{Historical Audio Archive Integrity:} Ensuring the accuracy and integrity of historical recordings by identifying unauthorized edits or restorations. Applies to  `Equalization' \id{equalization}, `Noise Reduce' \id{noise_reduce} and `Overlay Background Noise' \id{overlay_background}.
\end{itemize}

\section{Conclusion}

In this paper, we argue that the conventional binary classification of voice recordings solely into `fake' and `benign' categories, particularly in the context of Text-to-Speech (TTS), oversimplifies the complexity of the issue. 
A set of low-level edits can been utilized to disseminate misinformation, undermine political figures, and perpetrate fraud. 
The identification of such voice edits is therefore critical. 
We introduce a detailed taxonomy of voice edits, compile a dataset framework corresponding to these categories, and assess the efficacy of existing anti-spoofing and deepfake detection models. 
Most importantly, we argue for a paradigm shift in research priorities: 
moving beyond the mere identification of AI-generated modifications to a more comprehensive examination that spans the full array of vocal edits.


\section{Acknowledgement} 
This work has been (partially) funded by the Bavarian Ministry of Economic Affairs, Regional Development and Energy; and the Department of Artificial Intelligence, Wrocław University of Science and Technology.

\bibliographystyle{IEEEtran}
\bibliography{mybib}

\end{document}